\documentclass[manuscript,screen,nonacm]{acmart}
\AtBeginDocument{%
  \providecommand\BibTeX{{%
    \normalfont B\kern-0.5em{\scshape i\kern-0.25em b}\kern-0.8em\TeX}}}

\setcopyright{acmcopyright}
\copyrightyear{2023}
\acmYear{2023}
\acmDOI{XXXXXXX.XXXXXXX}

\acmConference[MuC]{Make sure to enter the correct
  conference title from your rights confirmation emai}{September 03-06,
  2023}{Rapperswil, Switzerland}
\acmBooktitle{Mensch und Computer 2023} 
\acmPrice{15.00}
\acmISBN{978-1-4503-XXXX-X/18/06}

\begin{document}

\title[To Classify is to Interpret: Building Taxonomies from Heterogeneous Data]{To Classify is to Interpret: Building Taxonomies from Heterogeneous Data through Human-AI Collaboration}

\author{Sebastian Meier}
\email{sebastian.meier@fh-potsdam.de}
\orcid{0000-0002-7466-4059}
\affiliation{%
  \institution{Potsdam University of Applied Sciences, Interaction Design Lab}
  \streetaddress{Kiepenheuerallee 5}
  \city{Potsdam}
  \state{Brandenburg}
  \country{Germany}
  \postcode{14469}
}

\author{Katrin Glinka}
\email{katrin.glinka@fu-berlin.de}
\orcid{0000-0002-4232-8907}
\affiliation{%
  \institution{Freie Universität Berlin}
  \city{Berlin}
  \country{Germany}}

\renewcommand{\shortauthors}{Meier \& Glinka}

\begin{abstract}
Taxonomy building is a task that requires interpreting and classifying data within a given frame of reference, which comes to play in many areas of application that deal with knowledge and information organization. In this paper, we explore how taxonomy building can be supported with systems that integrate machine learning (ML). However, relying only on black-boxed ML-based systems to automate taxonomy building would sideline the users' expertise. We propose an approach that allows the user to iteratively take into account multiple model's outputs as part of their sensemaking process. We implemented our approach in two real-world use cases. The work is positioned in the context of HCI research that investigates the design of ML-based systems with an emphasis on enabling human-AI collaboration.
\end{abstract}




\maketitle

\section{Introduction}
\label{Introduction}
Human sensemaking and knowledge production are fundamentally dependent on classification~\cite{bowker_sorting_1999}. Taxonomies serve this purpose as structured classification schemes that adhere to domain-specific standards. The importance of organizing, segmenting, and classifying data is especially obvious in light of the ever growing amount of information that is being created, aggregated, and made available through specialized data repositories or on the Internet. In light of the amount and heterogeneity of the available data, classification can hardly be addressed by means of manual-cognitive processing alone. Systems that integrate machine learning (ML) are able to process large amounts of data and, thus, can help with the task of classification and organization. However, delegating this task to ML-based systems in their entirety would mean that we sideline human interpretation and rely on the output of black-boxed systems that reproduce language ideologies and representational harms (see, e.g.,~\cite{blodgett2020-languageBias}). As an attempt to highlight the interpretative character of classification and taxonomy building, we propose to design ML-based systems that enable human-AI collaboration. Such systems are designed with the goal to effectively combine human competencies and computational capabilities~(see, e.g.,\cite{Shneiderman_Maes_1997, Terveen_1995}). Our approach enables domain experts to iteratively interact with the suggestions of the system while retaining interpretative authority. We report on the concept and implementation of this approach that we realized for two real-world use cases.\footnote{The code of both use case is available under open source licenses. An interactive online demo with code instructions can be found here: https://github.com/sebastian-meier/muc23-human-ai-demo} In the following, we situate our work in the context of related research on human-AI collaboration.

\section{Related Work}

Although data and classifications make up a huge part of the infrastructures that mediate interactions, the creation of taxonomies has not been foregrounded in Human-Computer Interaction (HCI) research~\cite{feinberg2014}. Nonetheless, research from HCI informed our exploration of how we can design ML-based systems that support taxonomic classification. We propose to design systems for iterative taxonomic classification in such a way that ML-outputs are framed as suggestions that need to be scrutinized by human interpretation in order to contribute to sense making (see, e.g.,~\cite{Baumeretal:2020:Topicalizer}). Likewise, we seek to balance the agency distribution between ML-based systems and their users, instead of aiming at full automation~\cite{Jiang_serendipity}. In this regard, our approach is informed by research into human-AI collaboration that is concerned with conceptualizing and designing interactions between humans and ``AI'' in such a way that the complementary capacities of humans and technology contribute to achieving a shared goal~\cite{Shneiderman_Maes_1997, Terveen_1995, Wang_HHColltoHumanAIColl}. Instead of aiming for automation, we conceptualize taxonomy building as an interactive and interpretative practice that integrates human competencies, machine learning, and information visualization. Our approach foregrounds the interpretative aspect of data generation~\cite{FeinbergDesignPerspectiveData2017} - both on the side of the human as well as on the side of the computer. 

\section{Problem Space}
One important step in the processing and organization of data is classification. Classification is often done within a taxonomy and alongside domain-specific standards. Generally speaking, taxonomies are controlled vocabularies that are intertwined with data standardization practices. 
Our work addresses two common challenges that arise in this context. For one, we might be confronted with data that is being created dynamically and within a frame of reference that is not clearly pre-defined. Consequently, patterns of meaning that could be grouped under one or more classifiers only emerge over time. Additionally, in cases where such dynamically created data does not relate to a clearly defined frame of reference, we cannot base our classification on already existing taxonomies.
We address this challenge with a use case on the classification of crowd-sourced questions submitted to a citizen-driven science agenda setting project in~\autoref{UseCaseI}. 
For another, even when working within an already established taxonomy, data work is often done manually by humans. This means that the data created might not strictly follow the established taxonomy or classification system. As Bowker and Star have pointed out 
, there is ``a permanent tension between attempts at universal standardization of lists and the local circumstances of those using them''~\cite[p.139]{bowker_sorting_1999}, which leads to modification of the classification system. Even in cases where taxonomies serve the needs of those who use them, errors can occur. 
Thus, the existence of a taxonomy alone does not guarantee consistent and clean data. We address this challenge with a use case on the taxonomic classification of open government data as part of an attempt of homogenizing the metadata in a step towards functional linked data \cite{bernerslee_2006} 
in~\autoref{UseCaseII}. 


\section{Use Cases}
In the following, we briefly describe both use cases in terms of context, intended user groups, underlying dataset and purpose of taxonomic classification. In both implementations, our approach aims at making large datasets usable (e.g. explore, search, find) for 
users that do not have a background in data science or ML. 

\subsection{Use Case I - Crowd-sourced Questions}
\label{UseCaseI}

Germany's Federal Ministry of Education and Research (BMBF) planned to deploy a large-scale participation project in 2022 (see~\cite{bmbf2022}) to develop a citizen-driven science agenda informed by the citizen submitted questions. The first use case was an initial prototype developed as a testbed for the future platform.
A similar project had been launched by the Research Foundation Flanders (FWO) in Belgium in 2018. The FWO project amassed 10.559 questions (see~\cite{vraagvoordewetenschap}). Our prototype aimed at helping the project managers of the BMBF to organize the submitted questions by way of taxonomic classification. The resulting taxonomy should, for one, provide the Federal Ministry with an overview of the submitted questions. For another, it should allow citizens to explore the already submitted questions and provide them with a frame of reference for their own questions during the submission process. We used the questions from the FWO survey as a sample dataset (see \cite{vraagvoordewetenschap2}). 
While the overall frame of reference is `science', citizens were free to ask any questions that they believe science can (or should be able to) answer. Therefore, the result is a heterogeneous dataset in regards to language (choice of words, style of writing, etc.) as well as content and length (from 9 characters to 274). One could try to impose academic taxonomies (i.e., relating to different fields of science) onto the questions. However, this would disregard the citizens' perspective and expertise. In contrast, our system encourages the building of a new taxonomy based on the citizen's questions. In summary, the system's intended users are project managers of the Federal Ministry. The dataset in question is dynamically generated by citizens. In this regard, the full extent of the dataset and its content are unknown to the users.

\subsection{Use Case II - Open Government Data - Metadata}
\label{UseCaseII}
Open government data (OGD) has become a cornerstone of government transparency around the world. To organize the datasets, metadata standards have been created by European and German working groups (e.g., INSPIRE~\cite{INSPIRE} and DCAT-AP~\cite{DCAT}). Each of these metadata standards contain controlled vocabularies to classify each dataset's theme or category\footnote{INSPIRE themes~\cite{INSPIRE_themes} (level 1) and codes~\cite{INSPIRE_codelist} (level 2), DCAT-AP:  Data/MDR themes~\cite{DCAT_themes}}. Additionally, datasets can be described through keywords and tags, for which no pre-defined taxonomy exists. As a result of the hierarchical structure of the federal government in Germany, OGD is collected and provided through various bodies within this structure (e.g. federal states, municipalties and government institutions) and distributed across various organizational and technological infrastructures. While guidelines for using the mentioned standards exist, they are not enforced and data is often not validated. Therefore, the `theme' attributes are rarely used correctly (left empty, false entries, typing errors, etc.). The result is a very heterogeneous dataset. For our use case, we took all the data that could not be linked with the official taxonomies and combined the `themes' entries with the `keywords' entries (as for most cases the themes field was empty). This resulted in 154.605 descriptors. To reduce the number of unique descriptors, we pre-processed the data in five steps: 1) removing items with only numbers or dates, 2) removing items that only contain stop-words, 3) removing items matching OpenStreetMap place names\footnote{Although place names seem like a useful descriptor, they are not allowed as themes for datasets within the established metadata standards.} 4) merging of small typos through a) removing space surrounding descriptors, b) conversion to all lower-case characters and c) Levenshtein distance and fingerprinting measures (inspired by OpenRefine \cite{OpenRefine}), 5) removal of descriptors used only three times or less. This resulted in 16.207 presumably unique descriptors. This second use case aims at enabling the intended users (ODG experts) to explore the descriptors and re-classify them in order to build a taxonomy that conforms to the existing metadata standards. The intended user group for this use case are OGD experts who have no expert knowledge of ML.

\section{Proposed approach}
Generally speaking, building and applying a taxonomy for a dataset involves trying to find commonalities between items. Identifying commonalities hinges on first establishing a guiding (and meaningful) principle for the assessment of `similarity' between items. In both of our use cases, we only work with text-based items. For this purpose, we deem text-based content similarity a meaningful principle of similarity. A variety of NLP and ML approaches exist that enable this type of content similarity analysis (see~\autoref{EmbeddingVis}). General purpose models make NLP readily applicable to a range of tasks, even when there is no domain-specific model. However, this also means that the text-to-vector embeddings will contain a level of uncertainty. Automatically building a taxonomy based on such imperfect classifications made by a black-boxed ML model would not emphasize the interpretative aspect of data generation and taxonomy building, which we consider to be a crucial element of human sensemaking and knowledge production. Furthermore, even with general purpose models, different versions exist in which certain parameters can be adjusted. In the following, we describe how our proposed approach frames the models' output as suggestions that the user can interact with. In regards to use case II, we also illustrate how our approach foregrounds the impact of different model parameters on the system's output. We lay out the technological and design elements of our system in reference to existing approaches in the following.

\subsection{Embedding Visualization Systems}
\label{EmbeddingVis}
\begin{figure*}[h]
  \centering
  \includegraphics[width=0.75\linewidth]{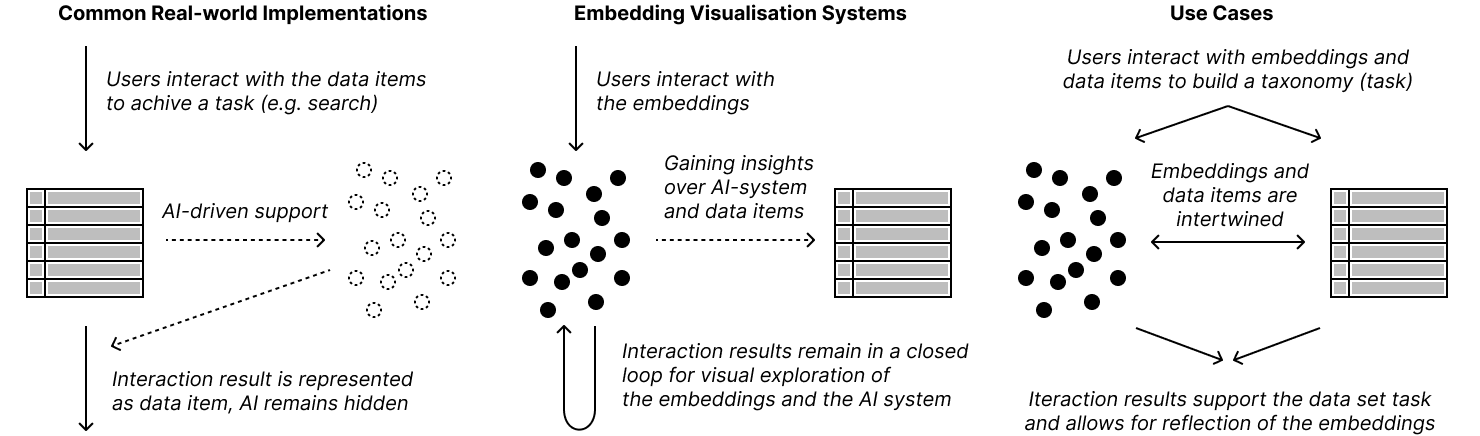}
  \caption{Interactions with text-to-vector embeddings in common implementations, embedding visualisation systems, and our use cases}
  \label{fig:related_works}
  \Description{The graphic compares the interactions with text-to-vector embeddings in a) common implementations, b) embedding visualisation systems, and c) the use cases described in this paper. For all three interaction sets, the graphic uses arrows to highlight the direction of interactions. In the first case - the common real world implementations - the users interact with the data item to achieve a task (e.g., search). To do so, the users interact with an abstracted visual user interface. The AI-support (e.g., based on text-to-vector embeddings) that leads to the final interaction result is hidden underneath the user interface. In the second case - the embedding visualisation system - the users interact with visualizations of the embeddings of the data items. Users gain insights into the data items or rather their embeddings (e.g. clusters) through interaction with the visualization. The interactions are usually not connected with a third system. Therefore, the interaction and insights stay in a closed loop for visual exploration of embeddings. In the last case - the use cases described in this paper - the users interact with the the data items and their embeddings and are enabled to build a taxonomy for the data items. Data items and their embeddings are visually connected. Thereby, the AI-generated embeddings help in the actual user task (taxonomy building) while remaining visible (unlike the first case).}
\end{figure*}

Interactions with text-to-vector embeddings can be found in many applications (e.g. search engines or recommender systems highlighting related items). In those systems, the user only interacts with the data items, while the underlying embeddings are hidden (see~\autoref{fig:related_works} left). Alternatively, embedding visualizations allow users to explore the embeddings visually, search related items, or to use clustering algorithms, just to name a few. Although not usually present in common real-world implementations, such visualisations of embeddings are not a novel approach. Existing examples include Google's embedding projector~\cite{TensorflowGithub} or Ji et al.'s visual analysis tool of cluster methods~\cite{8667702}. 
However, existing visualization approaches and systems do not connect the interactions with the embeddings in such a way that the user can perform data tasks (i.e. alter, adjust, or re-assign classifications) on the underlying dataset (see~\autoref{fig:related_works} middle). 

In contrast, our system enables interaction with the embedding visualizations to structure and organize data items, i.e., the embeddings and their visualisation are used to perform a task on the dataset itself (see~\autoref{fig:related_works} right). This way, visualizations not only provide users with an overview of what the dataset is about, but also constitute a crucial component in the workflow and sensemaking process. 
The embedding visualizations are not a monolithic and static component of the interface, but rather a canvas on which the users can iteratively and interactively build a taxonomy - guided by their own interpretation of the system's suggestions that they can relate to the underlying texts or entries.

\subsection{Data-Processing Pipeline and Machine Learning}

In both use cases we calculate the similarity of 1) short paragraphs (use case I) and 2) category descriptors (use case II), of one to n-words length, with text-to-vector approaches. The default model used is TensorFlow's universal-sentence-encoder v4 (USE) (see~\cite{DBLP:journals/corr/abs-1803-11175, TFHUB_use}). However, since the data from our use cases is in Dutch (use case I) and German (use case II), we added translation into English to the pipeline ~\cite{GoogleTranslate}. While translation adds to the uncertainty of the pipeline, the improved precision and performance of USE for English justified this trade-off. 

With use case II, which focuses on OGD experts as user groups, we decided to allow the users to interact with more models and added USE ``Large'', Standard English and Multilingual \cite{yang2019multilingual} as well as Neural-Net Language Models (NNLM) for English and German \cite{10.5555/944919.944966}\cite{TFHUB_nnlm}. The result is a 512-(USE) or 128-dimensional (NNLM) vector. 

For the 2D visualisation, we applied dimensionality reduction methods to the vectors (use case I: T-SNE \cite{van2008visualizing}), use case II: T-SNE, PCA and MDS). Users can  furthermore alter dimensionality reduction parameters. For the expert users in use case II, we also applied the DBSCAN \cite{10.5555/3001460.3001507} clustering algorithm in order to provide visual cues that would help them to identify potential clusters within the individual embedding sets. The reason behind providing this range of models and parameters in use case II is that each step, especially the dimensionality reduction, introduces uncertainty and variability. This way, the variety and inconclusiveness of ML-based similarity analysis is made visible to the users.

To achieve fast interactions, we ran all the above steps in advance, which generated 448 different model outputs. To give users in both use cases an additional method to identify similar items, we are using a VPTree \cite{10.5555/313559.313789} implementation to allow users to quickly browse the full embeddings.

\subsection{Interaction Flow Concept and Interface Design}


\begin{figure*}[h]
  \centering
  \includegraphics[width=0.75\linewidth]{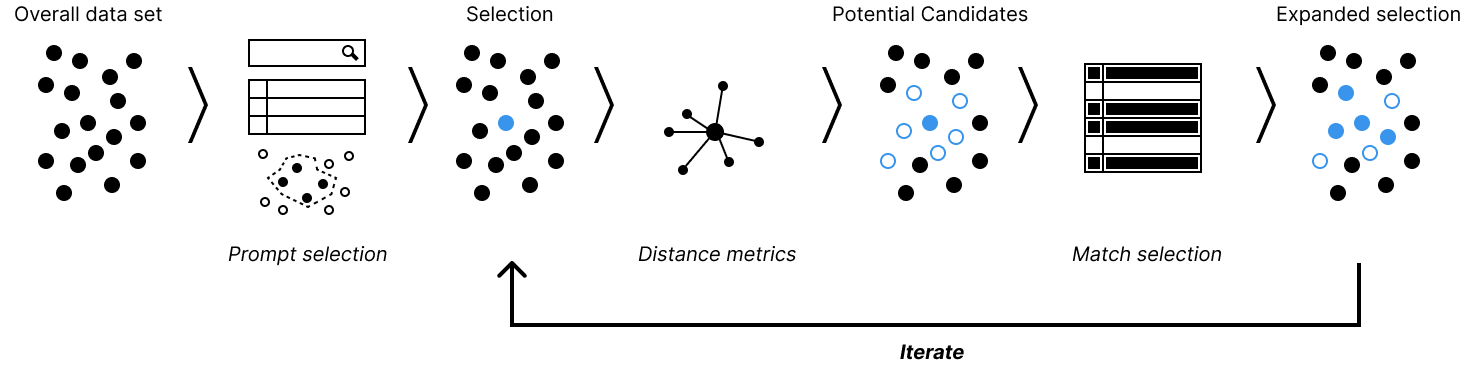}
  \caption{Iterative selection process, based on similarity measures on the text-to-vector embeddings}
  \label{fig:iterative}
  \Description{The graphic shows the process of the iterative selection process, based on similarity measures on the text-to-vector embeddings. From left to right the steps are described. 1. the overall data set is shown, 2. users can select a starting point from either the embeddings or a text search of the data items, 3. the result is the initial selection of items, 4. based on this selection, similar items (based on embeddings) are shown to the user as potential candidates, 5. the user can then select good candidates and remove bad ones. This new selection is then used as a new starting point for repeating the steps 4 or 5, in order to turn this process into an incrementally improving and growing selection process.}
\end{figure*}

While the underlying data processing- and NLP-pipelines are the same, the interactive graphical user interfaces (GUIs) explore two distinct directions, 1) a simplified interactive GUI for the project managers (use case I), and 2) a more advanced interactive GUI for OGD-experts (use case II) (see~\autoref{Appendix} for annotated screenshots of both GUIs).

The interaction flow concept of both GUIs is focused on an iterative process that refines the selection of potentially similar items to be considered in the taxonomic classification (see~\autoref{fig:iterative}). Starting with a data item selected by the user, potentially similar candidates (based on the VPTree) are suggested to the user in list and point cloud form. This allows the user to select the best candidates for the new taxonomy group. Beyond VPTree, users can draw a polygon-hull to select points from the visualisation. This then increases the list of potential candidates. As more and more manually matched candidates are selected for the cluster, the search for additional potential candidates is expanded. The users can also remove candidates and add them on a `ignore' list, which will be taken into account by subsequent searches. 

\begin{figure*}[h]
  \centering
  \includegraphics[width=0.75\linewidth]{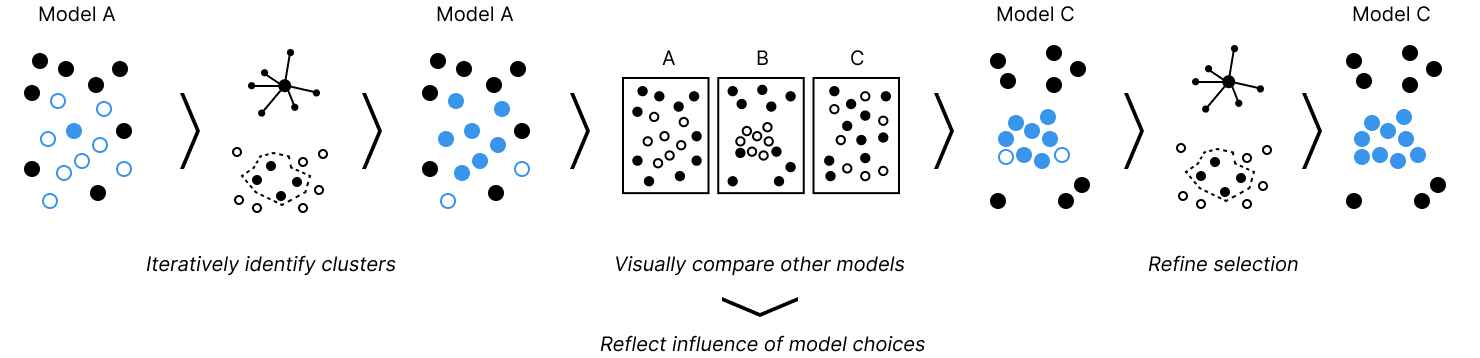}
  \caption{Extension of fig. \ref{fig:iterative}, by integrating multiple embeddings and enabling users to compare and switch between models}
  \label{fig:comparison}
  \Description{The graphic extends the previous process of the iterative selections. In the previous example, users could refine their selection through suggestions of the text-to-vector embeddings, but the embeddings were fixed. In other words: they were based on one specific embedding process. In the extension of this process in use case II, which is illustrated in the graphic, the users can switch between various embeddings for the data set. Depending on the selected model, different patterns of similarity emerge. Users can then visually compare the output of different models and reflect on the influence of model choice.}
\end{figure*}

For the second use case, the users can visually analyse the current selection of candidates across multiple models' embeddings (see~\autoref{EmbeddingVis}). Thereby, we allow users to visually search for potential clusters in different settings (see~\autoref{fig:comparison}). By integrating multiple embeddings, we also believe to work towards increasing awareness for the biases, uncertainty and variability in model outputs by emphasizing embeddings as suggestions - suggestions that ultimately need human interpretation to successfully contribute to taxonomic classification of heterogeneous data.

\section{Discussion}
\label{sec:discussion}

The two use cases provided us with a frame of reference for the development of two prototypes through which we explore the design space for taxonomic classification through human-AI collaboration.
Usually, in applications with high levels of automation, the user is only presented with one outcome, which obfuscates how much the prediction depends on model configuration. In such cases, the model's output acts as a normative statement. The problems with overreliance in AI and automation bias are well known and are being addressed, for example, in the context of research into explainable AI (XAI)~\cite{BuccinaCFF}. However, research also suggests that explanations might even have a worsening effect, in the sense that offering explanations could lead to even more overreliance and trust in AI~\cite{BansalXAITeamPerformance}. We propose to take an alternative approach that tries to balance levels of automation (see, e.g.,~\cite{mackeprang_sweet}) and combine it with an emphasis on the variability of model outputs. With this, we seek to frame the model's output as an interpretative statement. We do so by integrating faceted linked views, small multiples, and visual analytics and, thereby, make the embedding visualisations part of the interface instead of hiding them underneath.

\section{Limitations and Future Work}
As a next step we have to evaluate our approach through user studies in order to confirm whether and to what extent the intended effects also hold true in a user study. While we acknowledge this obvious limitation, we consider the work presented in this paper as a starting point for further research. With these studies we seek to investigate if the facilitation of user reflection regarding the interpretative aspect of taxonomy building, were realized.

In regards to technological and design limitations, we had to take a pragmatic approach towards the selection of models and parameters. Since use case I is focused on project managers of the BMBF, we decided to simplify the GUI. This, in turn, counteracts our intention to confront the users with a variety of system outputs. In use case II, however, we offer 448 model outputs, which is too much to continuously visually compare. This bares the question of how to select the ``best'' models for the users to choose from, i.e., how to balance visual overload on the one hand while enabling visual exploratory analysis on the other hand.

After refining our approach, we plan to transfer it to other domains (e.g. digital cultural heritage). We believe that our approach can contribute to the effort of data organization while at the same time raising awareness for the variability and uncertainty of ML model outputs, the role of interpretation and importance of human sensemaking. 
\begin{acks}
The work presented in this paper was made possible through funding by the Federal Ministry of Education and Research (BMBF) and the Federal Ministry for Economic Affairs and Climate Action (BMWK).
\end{acks}

\bibliographystyle{ACM-Reference-Format}
\bibliography{sample-base}

\clearpage
\appendix
\section{Appendix: Annotated Screenshots}
\label{Appendix}

\begin{figure*}[h]
  \centering
  \includegraphics[width=\linewidth]{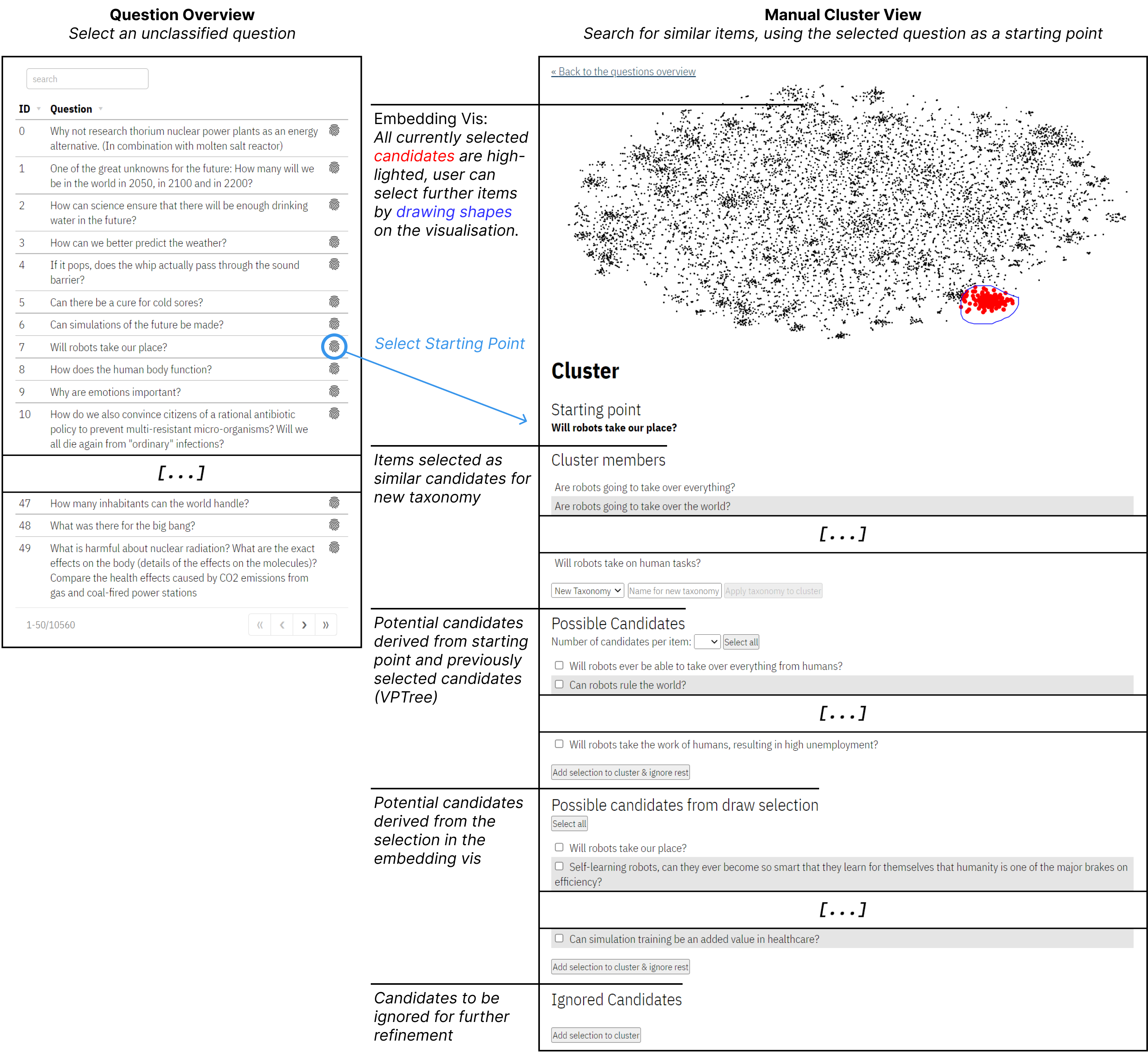}
  \caption{Annotated screenshot from use case I of the \textbf{question overview} (left) for selecting the starting point for the construction of a new class and the \textbf{manual cluster view} (right) for iteratively selecting class candidates}
  \label{fig:case-study-1}
  \Description{The figure shows two screenshots and corresponding explanations of the first prototype for the iterative taxonomy building (use case I). The interface design is minimalist using mostly black, white and gray tones, red is used as a highlight color. On the left side, a paginated list of question with an additional search slot allows users to navigate the data set of questions and select one of the questions as a prompt to start the taxonomy building process. From this ``starting point'' the user switches to the second sceen (displayed on the right side of the graphic). This screen shows the actual interactive interface. It starts with a typical TSNE scatterplot that shows all data points. The starting point is highlighted in red. Users can select additional candidates of `similar' items from this scatter plot or draw shapes around additional points to be included as potential candidates for the new taxonomy. The rest of the screen is split into four parts (organized in list form): 1. ``Cluster members'': include items selected for the new taxonomy, 2. ``Possible Candidates'': include suggestions from the distance analysis within the embeddings, 3. ``Possible candidates from draw selection'': include candidates selected through the scatter plot interactions, 4. ``Ignored candidates'': include candidates that the user excluded from one of the candidate lists. Each list is accompanied by buttons allowing the user to move items between those lists. Thereby, users can iteratively refine the suggestions by the system.}
\end{figure*}

\begin{figure*}[h]
  \centering
  \includegraphics[width=\linewidth]{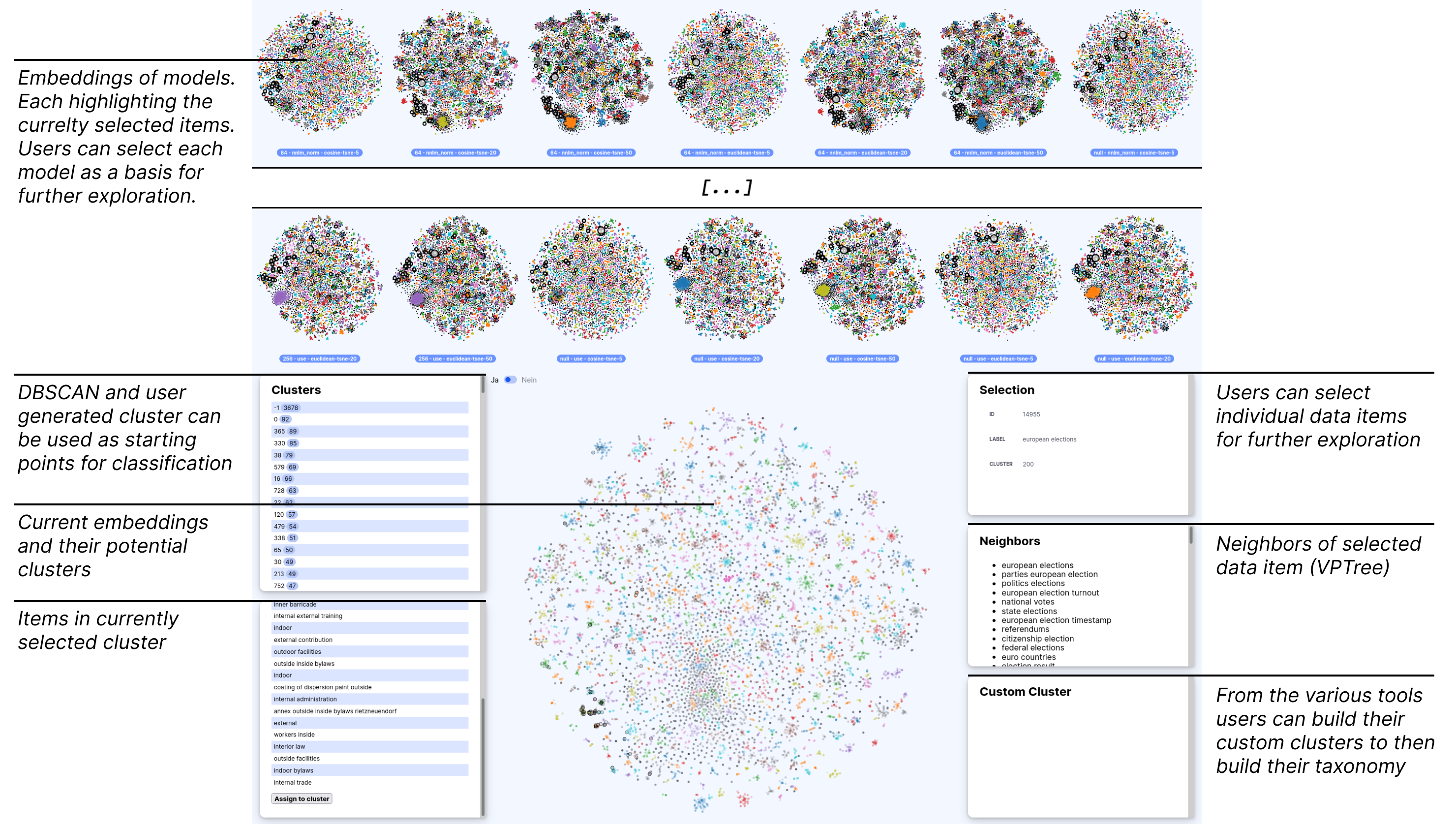}
  \caption{Annotated screenshot of the expert interface for use case II}
  \label{fig:case-study-2}
  \Description{The figure shows a screenshot with explanations for the various elements of the user interface. The interface is also presented in a minimalist design, using three shades of light blue, as well as black and white. Interactive buttons are presented in grey. The interface consist of an upper and lower area. In the upper area, 14 circular scatter plots, typical for TSNE visualisations, are shown in two rows that each contain 7 ``miniature'' scatter plots. The points in the plots are coloured based on an automated clustering process. Three dots between the first and second row of 7 scatter plots indicate that in the full interface there would be even more plots that the user can chose from. Each plot represents one embedding process, a label underneath each plot describes the model (e.g. universal sentence encoder, multi language, cosine, 128 output vectors). The lower part shows a big TSNE plot in the middle (which was selected from one of the ``miniature'' scatter plots above). This big plot is surrounded by interactive panels. Starting with a panel of system-generated clusters, each cluster has an id and a color, corresponding to the currently selected model in the middle. Below the cluster list is the currently selected cluster and the data items belonging to the class. The items within this selected cluster are highlighted on the big as well as on the 
 ``miniature'' plots, allowing the user to compare the models. The next panel shows more information (all attributes and id) on an individual item selected by the user. Below are the neighbour items to that item displayed. The last panel shows new clusters that were built by the user. To build new clusters, each panel includes buttons that allow the user to select all or individual items from a custom selection from the plot, neighbours or existing clusters.}
\end{figure*}

\end{document}